\documentclass[namedreferences]{SolarPhysics}
\usepackage[optionalrh]{spr-sola-addons} % For Solar Physics 
\usepackage{epsfig}                     % For eps figures, old commands
\usepackage{graphicx}                    % For eps figures, newer & more powerfull
\usepackage{color}                       % For color text: \color command
\usepackage{url}                         % For breaking URLs easily trough lines
\begin{document}

\begin{article}

\begin{opening}

\title{Solar-Terrestrial Simulation in the STEREO Era: The January 24-25, 2007 Eruptions}

%%%%%%%%%%%%%%%%%%%%%%%%%%%%%%%%%%%%%%%%%%%%%%%%%%%
%% Authors Names
%
\author{N.~\surname{Lugaz}$^{1}$\sep
        A.~\surname{Vourlidas}$^{2}$\sep
        I.~I.~\surname{Roussev}$^{1}$\sep
        H.~\surname{Morgan}$^{1,3}$      
       }

%%%%%%%%%%%%%%%%%%%%%%%%%%%%%%%%%%%%%%%%%%%%%%%%%%%
%% Runningheads
%
\runningauthor{Lugaz et al.}
\runningtitle{Jan. 24-25, 2007 CMEs}

%%%%%%%%%%%%%%%%%%%%%%%%%%%%%%%%%%%%%%%%%%%%%%%%%%%
%% Affilations 
%
  \institute{$^{1}$ Institute for Astronomy - University of Hawaii-Manoa, 2680 Woodlawn Dr., Honolulu, HI 96822
                     email: \url{nlugaz@ifa.hawaii.edu}, \url{iroussev@ifa.hawaii.edu} , \url{hmorgan@ifa.hawaii.edu}\\ 
             $^{2}$ Code 7663, Naval Research Laboratory, Washington, DC 20375
                     email: \url{vourlidas@nrl.navy.mil} \\
             $^{3}$ Sefydliad Mathemateg a Ffiseg, Prifysgol Aberystwyth, Ceredigion, Cymru, SY23 3BZ 
             }

%%%%%%%%%%%%%%%%%%%%%%%%%%%%%%%%%%%%%%%%%%%%%%%%%%%
%%% Abstract 
\begin{abstract}
The SECCHI instruments aboard the recently launched STEREO spacecraft enable for the first time the continuous tracking of coronal mass ejections (CMEs) from the Sun to 1~AU.
We analyze line-of-sight observations of the January 24-25, 2007 CMEs and fill the 20-hour gap in SECCHI coverage in January 25 by performing a numerical simulation using a  three-dimensional magneto-hydrodynamic (MHD) code, the Space Weather Modeling Framework (SWMF). We show how the observations reflect the interaction of the two successive CMEs with each other and with the structured solar wind.
We make a detailed comparison between the observations and synthetic images from our model, including time-elongation maps for several position angles. Having numerical simulations to disentangle observational from physical effects, we are able to study the three-dimensional nature of the ejections and their evolution in the inner heliosphere. This study reflects the start of a new era where, on one hand, models of CME propagation and interaction can be fully tested by using heliospheric observations and, on the other hand, observations can be better interpreted by using global numerical models.
\end{abstract}

%%%%%%%%%%%%%%%%%%%%%%%%%%%%%%%%%%%%%%%%%%%%%%%%%%%
%% Keywords
%
%\keywords{}

\end{opening}
%-------------------------------------------------

%%%%%%%%%%%%%%%%%%%%%%%%%%%%%%%%%%%%%%%%%%%%%%%%%%%
%% Sections
%
 \section{Introduction}\label{sec:Intro} 
While Coronal Mass Ejections (CMEs) have been observed and studied extensively since the late 1960s (see reviews by \citeauthor{Schwenn:2006}\citeyear{Schwenn:2006} and \opencite{Roussev:2006}), there had been no continuous dedicated remote-sensing observations of the heliosphere between 0.15~AU and 1~AU from the end of the {\it Helios} missions in the early 1980s to the launches of the {\it Solar Mass Ejection Imager (SMEI)} and the {\it Solar Terrestrial
Relations Observatory (STEREO)} in 2003 and 2006, respectively. The in-depth study of the evolution of CMEs in the heliosphere had to rely solely on numerical models, mostly for ``generic'' fast CMEs (e.g., see \opencite{Odstrcil:1999} and \citeauthor{Manchester:2004a}\citeyear{Manchester:2004a}). Only in the past two years there have been Sun-to-Earth numerical simulations of real events (e.g., \citeauthor{Chane:2006}2006, \citeauthor{Lugaz:2007}2007 and \citeauthor{Toth:2007}2007). 
Particular attention has been given to the numerical modeling of CMEs at or near solar minimum, especially the May 12, 1997 CME \cite{Odstrcil:2004, WuCC:2007, Cohen:2008a, Titov:2008}. One reason for this focus is that the solar wind is believed to be simpler and steadier during solar minimum than during solar maximum. Therefore, solar minimum is thought to be the perfect time period to study the evolution of isolated CMEs. However, these previous works have not been totally successful in reproducing the observed transit time and the measured plasma parameters at 1~AU. Possible reasons are that the initial speed of the CME is not well constrained because Earth-directed CMEs appear as halo events and that there are no observations between 32~$R_\odot$ and Earth. Therefore, studying limb CMEs with remote heliospheric observations should help better constrain the existing solar wind and eruption models.

The two events on January 24 and 25, 2007 were the first fast CMEs observed by the Sun Earth Connection Coronal
and Heliospheric Investigation (SECCHI) suite \cite{Howard:2008}. Observations by the Heliospheric Imagers have been reported in \citeauthor{Harrison:2008}2008 and numerical simulations have been performed by a number of groups and reported in publications \cite{Lugaz:2008b} and at meetings (e.g., \citeauthor{Webb:2008}2008). In contrast to the May 12, 1997, this was not an isolated event but two successive CMEs which are believed to have interacted on their way to 1~AU. Interacting CMEs are not uncommon, especially near solar maximum \cite{Gopalswamy:2002} and there have been recent efforts to model them numerically \cite{Odstrcil:2003, Lugaz:2005b, Lugaz:2007, Xiong:2006}. Even though CME-CME interactions are less frequent near solar minimum, there have been few reported cases, notably the April 11, 1997 geomagnetic storm, which is thought to have been caused by two CMEs \cite{Berdichevsky:1998}.

Studying the January 24-25, 2007 CMEs is important for future space weather predictions, because interacting CMEs are a major cause of large geomagnetic storms \cite{Zhang:2007}.  It is also the first time that we can track this type of events in the heliosphere with actual observations. However, there are a few downsides: these were limb CMEs, therefore there were no in situ observations at Earth (nor by the STEREO spacecraft). Also, limb events (with respect to the observing spacecraft), such as these CMEs, may become undetectable by SECCHI/Heliospheric Imagers as well as SMEI at large elongation angles \cite{Vourlidas:2006}. In addition, the passage of comet McNaught increased the background emission significantly in the south-eastern quadrant of the observations at the time of the eruptions. Last, but most importantly, there was a 20-hour data gap in the SECCHI coverage, gap, which coincided with the likely time of the interaction between the ejections. 

Detailed studies of synthetic line-of-sight images have been performed by a number of groups in the past few years \cite{Chen:2003,Manchester:2004a, Lugaz:2005a, Odstrcil:2005,Riley:2008}. Comparison of such synthetic observations with real observations by the Large Angle and Spectrometric Coronagraph experiment (LASCO), SMEI or STEREO has only been done in the past three years for a few selected events during or close to solar maximum \cite{Lugaz:2007, Manchester:2008, Sun:2008}.

In this article, we report our efforts to understand the January 24-25, 2007 coronal mass ejections by means of three-dimensional (3-D) magneto-hydro-dynamic (MHD) simulations using the Space Weather Modeling Framework (SWMF) developed at the University of Michigan \cite{Toth:2005}. In Section \ref{sec:Obs}, we discuss the observations as well as the simulation set-up. In Section \ref{sec:Sim}, we discuss the evolution of the two CMEs in the heliosphere and their interactions, before comparing synthetic line-of-sight images with real ones, including time-elongation plots in Section \ref{sec:Comp}. We conclude and discuss the main findings of our investigation and their consequences for future observations of CMEs by STEREO in Section \ref{sec:Conclusion}.

%%%%%%%%%%%%%%%%%%%%%%%%%%%%%%%%%%%%%%%%%%%%%%%%%%%%%%%%%%%%%%%%%%%%%%%%%%%%
\section{The Solar Eruptions of 2007 January 24-25} \label{sec:Obs}
%%%%%%%%%%%%%%%%%%%%%%%%%%%%%%%%%%%%%%%%%%%%%%%%%%%%%%%%%%%%%%%%%%%%%%%%%%%%%

\subsection{LASCO and SECCHI Coronagraphs and Imagers}

LASCO onboard the {\it Solar and Heliospheric Observatory}  (SoHO) has two working coronagraphs observing the solar corona from 2.1 to about 32~$R_\odot$ with a cadence of about 40 minutes \cite{Brueckner:1995}. 
The coronagraphs and heliospheric imagers part of the SECCHI investigation onboard STEREO are COR-1, COR-2, Heliospheric Imager 1 and Heliospheric Imager 2 (HI-1 and HI-2, respectively). Their fields-of-view are 2.13$^\circ$ (4~$R_\odot$ with a 1.5~$R_\odot$ occulting disk), 8$^\circ$ (15~$R_\odot$ with a 2~$R_\odot$ occulting disk), 20$^\circ$ and 70$^\circ$ respectively. Also, the HIs are not pointed at the Sun but along the Sun-Earth line and their fields-of-view are offset by an angle of 13.65$^\circ$ and 53.35$^\circ$ with respect to the Sun-spacecraft line respectively \cite{Howard:2008}.
At the time of the ejections, STEREO was in its commissioning phase and STEREO-A was rolled by 22.4$^\circ$ from
the solar north, which resulted in HI-1 and HI-2 imaging higher-latitude
regions than during the normal phase of the mission \cite{Harrison:2008}.  
The spacecraft was separated by approximately 0.4$^\circ$
from Earth and was at a radial distance of 0.97~AU from the Sun.

\subsection{Coronagraphic and Heliospheric Observations}\label{LASCO}
On 2007 January 24, at 14:03UT, COR-1 observed an
eruption whose source region was behind the eastern limb of the Sun.  At 06:43UT on January 25, COR-1 observed a
second eruption also originating from behind the Sun. The similarities between these two eruptions, as observed by LASCO coronagraphs and SECCHI/CORs, lead us to conclude that they originated from the same source region. There were two active regions (ARs) which could be the source of these eruptions, both on the back side of the Sun: AR~940, which crossed the eastern limb in early January 26 and AR~941, which crossed it around mid-day on January 28. It is also possible that the two eruptions originated from different active regions, or that they may not be associated with any active region. However, we believe that the ejections originated from AR~940, 
%because there does not appear to be a large ``backside'' component 
because it had a more complex topology, and was the source of more activity than AR~941 as they were Earth-facing. More importantly, AR~940 was closer from the eastern limb: ARs~940 and 941 were about 24$^\circ$ and 43$^\circ$ behind the eastern limb, respectively, at the time of the first eruption. For all these reasons, we believe that AR~940 is the most likely source region for the studied limb CMEs. However, this is an assumption which may cause errors as large as 40$^\circ$ in the central position of the simulated ejections. Because the two STEREO spacecraft were not separated yet, it is not possible to identify the source region by triangulation as done by \inlinecite{THoward:2008b}. As they moved out of the field-of-view of SECCHI COR-1, these two eruptions were detected by LASCO C2 and C3 and by SECCHI COR-2.

The leading ejection was first detected in HI-1's field-of-view at 18:01UT on January 24
at 4.4$^\circ$ elongation and was tracked until 12.1$^\circ$ at 04:01UT
on January 25.  At that time, there was a SECCHI data gap for 20 hours.
Multiple fronts were detected in HI-2 on January 26: the brightest one
propagated from 24.7$^\circ$ at 02:01UT to 32.5$^\circ$ at 18:01UT.
\inlinecite{Harrison:2008} noted the presence of ``forerunner'' structures
(at 42$^\circ$ at 18:01UT). In our previous work about these events, we focused on the observations in HI-2 on January 26, 2007 \cite{Lugaz:2008b}. 
We showed that the two ejections had merged along most position angles (PAs) by the time SECCHI resumed operations. We also showed that the three brightest fronts observed in HI-2 during this day were associated with the two CMEs as well as a dense stream which had been compressed due to the passage of the CMEs. The brightest front was the second one, and it was associated, depending on the PA, with the merged CMEs and with the second CME propagating into the dense sheath of the first CME. 

\subsection{Simulation Set-up}
The 3-D simulation of the two ejections was done using the SWMF \cite{Toth:2005} as summarized in \inlinecite{Lugaz:2008b}.
The entire domain is a cube of (440~$R_\odot$)$^3$ centered on the Sun and divided into two sub-domains: the solar corona which is a cube of (48~$R_\odot$)$^3$ centered on the Sun and the inner heliosphere (IH) which fills the entire domain. The solar corona domain of SWMF is resolved with 40,489 blocks of $4^3$ cells each, ranging in size from
1/40~$R_\odot$ at the inner boundary to 0.75~$R_\odot$ at the outer
boundary.  
The IH is initially resolved with 16,626 blocks of $8^3$ cells each, ranging in size from 3.44 to
0.215~$R_\odot$ at its inner boundary of 16~$R_\odot$. 
The two domains are coupled as explained in \inlinecite{Toth:2005}
via a spherical buffer which spans between 16 and 17~$R_\odot$ in the two domains. 
In both domains, the numerical grid at the heliospheric current sheet is refined 
in order to better capture the density gradients there.
At later times, the grid in the IH is also refined during the merging of the two CMEs along the approximate location of the maximum scattering in the ecliptic plane (Thomson sphere) with cells as small as 0.215~$R_\odot$. 
%The maximum number of blocks in IH is 18,579 at 2300 UT on January 25; the total maximum number of cells for the entire simulation is %therefore about 12.1 millions. 

\subsection{Solar Wind and CME Models}

The solar wind and coronal magnetic field are simulated using
the model developed by \inlinecite{Cohen:2007}. This model makes use of
solar magnetogram data and the Wang-Sheeley-Arge (WSA) model
\cite{Wang:1990}. In this model, the polytropic index is modified in the corona based on the value of the velocity at 1~AU from the WSA model, so that the total energy is conserved over distance along field lines (Bernoulli integral).
This model has been validated  in previous studies of the evolution of a CME from the Sun to the Earth \cite{Cohen:2008a, Cohen:2008b}.
The solar magnetic field is reconstructed from a Legendre polynomial
expansion of order 49 based on NSO/SOLIS magnetogram data \footnote{obtained from the SOLIS website -- \url{http://solis.nso.edu}--}.
The initial coronal magnetic field is obtained through the potential field source surface model, before being let to relax to a non-potential steady state, obtained by solving self-consistently the MHD equations.  

To model the CMEs, we use a semi-circular flux rope prescribed
by a given total toroidal current, as in the model by \inlinecite{Titov:1999} implemented by \inlinecite{Roussev:2003b}. 
This model has been used in a number of previous studies of CMEs at solar minimum \cite{Cohen:2008a} and solar maximum \cite{Toth:2007} including the investigation of the merging of CMEs \cite{Lugaz:2007}. 
This flux rope solution, once superimposed on the background magnetic field, leads to immediate eruption
because of force imbalance with the ambient magnetic field. The flux rope parameters are chosen such that there is an agreement with
the observed values of the CME speed in the corona. The two flux ropes are initiated with the exact same parameters except for the value of the total current in the flux rope which is 75$\%$ larger for the second ejection than for the first one. The values used for the current (3.6 and 6.3 $\times 10^{11}$A) are in good agreement with typical values derived from coronagraphic observations of CMEs \cite{Subramanian:2009}.  
Finally, we added the flux ropes at 14:00UT on January 24 and 0640UT on January 25, in each case 3 minutes before the first observations by COR-1. The two CMEs are added onto the same position on the solar surface, therefore, they are separated by about 9$^\circ$ in longitude, as seen from a fixed point in the heliosphere. 

Note that this model is not aimed at reproducing the
complexity of the flux rope formation and the shearing motion at the
surface of the Sun as in the models of \inlinecite{Roussev:2007}, \inlinecite{Manchester:2008} or \inlinecite{Lynch:2008}.  
Our main goal here is to study the interaction of the two ejections with the
background solar wind and with each other, and to compare the observations made by LASCO and SECCHI with our simulation results. Since there was no observations of the evolution of the active regions prior or during the eruptions, a more complex model based on magnetic field observations cannot be used.

%% Figure 
 \begin{figure}[ht] 
 \centerline{\includegraphics*[width=12cm]{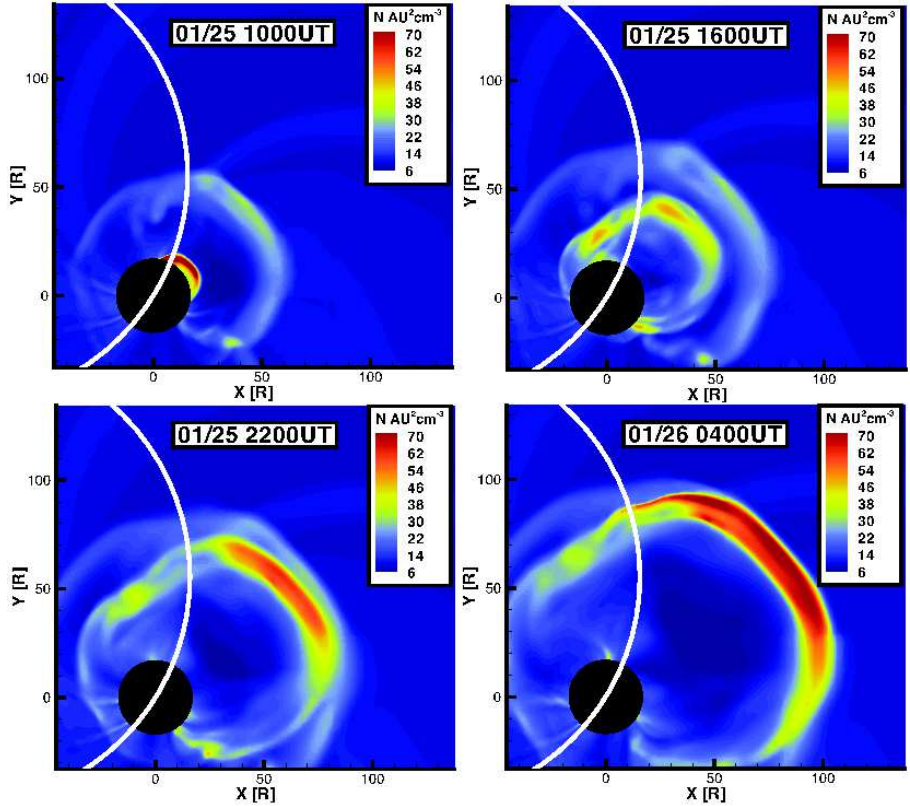}}
 \caption{View of the CMEs from the solar north at four different times prior and during their interaction showing the density scaled by
 1/R$^2$. The radius of the black disk 	is 16 $R_\odot$ and the white circle is the approximate projection of the Thomson sphere (relative to STEREO-A) onto the plane of the image. Earth's position is approximatively $(-179, 110, -20)R_{\odot}$ in this coordinate system.}
 \label{fig:2D}
\end{figure}

%%%%%%%%%%%%%%%%%%%%%%%%
\section{Evolution and Interaction of the Ejections in the Heliosphere: Filling the SECCHI Data Gap}\label{sec:Sim}
%%%%%%%%%%%%%%%%%%%%%%%%

At the launch of the second eruption, the first CME is about 55~$R_\odot$ away from the Sun with a radial speed of about 600~km~s$^{-1}$. The speed of the first CME decreases gradually to about 450~km~s$^{-1}$ over the course of the next day until it is overtaken by the second CME.
The second eruption is faster than the first one with a speed of about 1200 km~s$^{-1}$ at 20~$R_\odot$. First-order fits of the time-height profile as reported by the LASCO CME catalog for the two CMEs show that the second CME was 74$\%$ faster than the first one (1367 vs. 785 km~s$^{-1}$) in the upper corona. However, these speeds are for different position angles (PAs); %, and the speed of the first CME is derived for a ``protruding nose''-like feature near PA 55, which might be associated with a streamer ejection a few hours before the first eruption. I
if PA 90 is used to derive the speed for both CMEs, the difference in speed is greater (1360 vs. 600 km~s$^{-1}$). We, nonetheless, use a difference of 75$\%$ in the initial value of the electric current between the two CMEs as we have shown in previous studies that overtaking CMEs have less deceleration in the corona \cite{Lugaz:2005b} and we found this value to result in realistic velocities and deceleration in interplanetary space for the second CME. 

Because the two eruptions are more than 16 hours apart, the second eruption propagates into a solar wind which has essentially relaxed back to steady-state (as measured in density and velocity in the open field regions) after the passage of the preceding CME. 
%FIgure
 \begin{figure}[ht] 
 \centerline{\includegraphics*[width=10cm]{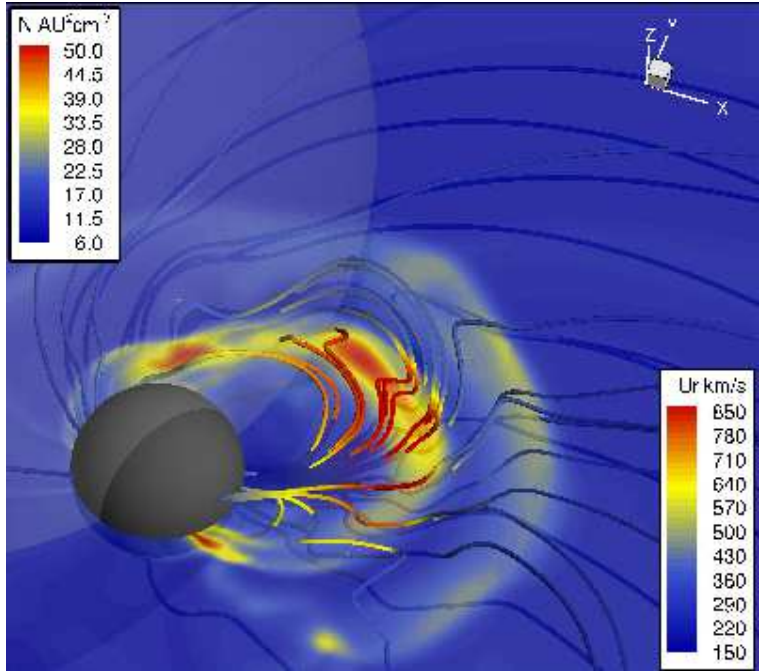}}
 \caption{View of the CMEs at 1600 UT on January 25 showing the density scaled by
 1/R$^2$ on the $z=0$ plane. The black sphere radius is 17~$R_\odot$ and the transparent sphere is the Thomson sphere (relative to STEREO-A). The 3-D magnetic field lines are color-coded with the radial velocity.}
 \label{fig:3D}
\end{figure}
%Figure%%%%%
This is different from our previous simulations of successive ejections \cite{Lugaz:2005b, Lugaz:2007} and show that, according to our solar wind model, the solar wind relaxation time is less than 16.5 hours after an ejection. The main difference between the state of the corona before the first and second eruptions is that two streamers on both side of the erupting active regions have been deflected by the first CME and have not come back to their pre-eruption positions. 

The second CME propagates into this quasi-steady solar wind ({\it top left panel} of Figure~\ref{fig:2D}) until approximatively 12:00UT on January 25 when it reaches the back of the first CME, characterized by faster speed and lower density, corresponding to the interplanetary manifestation of a magnetic cloud; the second CME front propagating into the low-density region associated with the first ejecta can be seen in {\it top right panel} of Figure~\ref{fig:2D}. 
There, the speed of the second CME decreases to about 1000~km~s$^{-1}$. The second CME then reaches the back of the dense sheath associated with the first CME around 18:00UT on January 25 and decelerates rapidly to speeds of about 800~km~s$^{-1}$; the second CME front propagating into the first CME front can be seen in {\it bottom left panel} of Figure~\ref{fig:2D}, and also in a three-dimensional view at the same time on Figure~\ref{fig:3D}. The two shocks have fully merged  by approximatively 04:00UT on January 26 ({\it bottom right panel} of Figure~\ref{fig:2D}). The evolution of these two ejections is overall similar to what has been described in great details in \inlinecite{Lugaz:2005b}. The speeds and positions in the description above are for the CMEs along the eastern limb (90$^\circ$ East of Earth); the CMEs are by nature 3-D and this is illustrated in Figure~\ref{fig:2D}, which shows the CMEs in the (solar) equatorial plane at the different stages of the interaction/merging, and Figure~\ref{fig:3D}, which shows a 3-D view corresponding to the {\it top right panel} of Figure~\ref{fig:2D}. 

According to our simulation, during the SECCHI down time (04:00UT to 23:59UT on January 25), the second ejection caught up with the first one, propagated into the lower density, faster material part of the first magnetic cloud and reached the dense sheath of the first ejection. However, we do not expect the eruptions to have yet fully interacted for all longitudes by the time SECCHI started imaging again (00:01UT on January 26). This is what we reported in \inlinecite{Lugaz:2008b}, identifying two of the bright fronts observed by SECCHI/HI-2 at 02:01UT on January 26 with the dense sheaths associated with the two CMEs. 

\section{Comparing Synthetic and Real Line-of-sight Observations} \label{sec:Comp}

\subsection{Coronagraphs and Heliospheric Imagers}
%Detailed comparisons between simulation results, synthetic coronagraphic images and LASCO images have been published before. In general, it has been shown that, for limb CMEs such as the January 24 and 25, 2007 CMEs, coronagraphs track the actual position of the CME sheath with high accuracy, and the speed derived from synthetic coronagraphic measurements is in good agreement with the three-dimensional speed of the ejecta. 

\begin{figure}[ht*] 
\centerline{\includegraphics*[width=12cm]{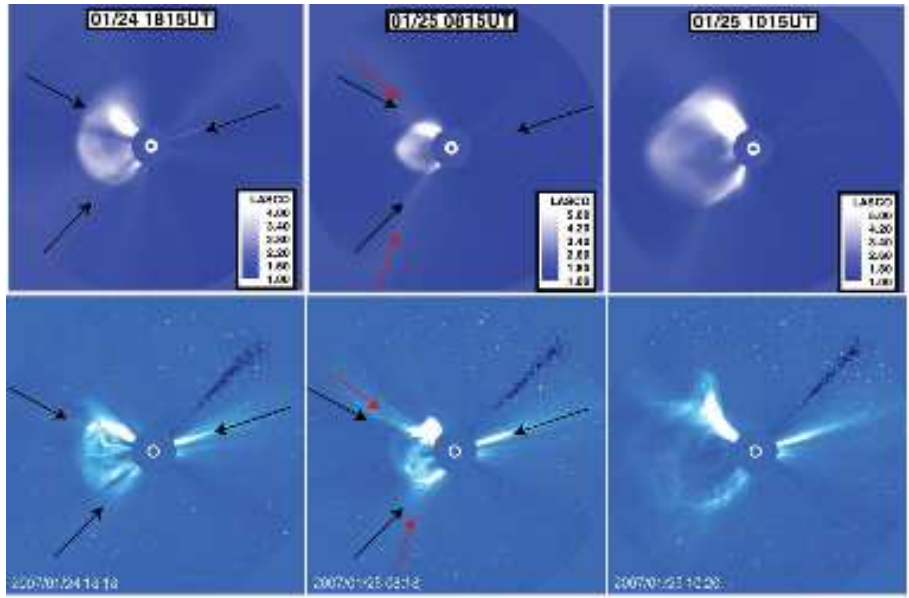}}
 \caption{Comparison of synthetic ({\it top}) and real ({\it bottom}) LASCO/C3 images at three different time instants, showing total brightness (divided by the monthly background for the synthetic images). Streamers' positions before the first ejection are shown approximatively with black arrows (same direction for synthetic and simulated images), the deflected streamers positions are indicated with the red arrows.}
 \label{fig:LASCO}
\end{figure}

In Figure~\ref{fig:LASCO}, we compare synthetic and real images of the two CMEs at three different times in the LASCO/C3 field-of-view. Note that, in this section, all images (synthetic or real) show total brightness. 
%%%%%%%%%%%%
%%%%%%%%%%%%%%%%
During late January 2007, there was a significant number of steady streamers (3 can be viewed in the LASCO images and their positions are indicated with black arrows). The two eastern streamers were disrupted by the two CMEs; their positions after the first eruption are marked with red arrows. We are able to capture the disruption of these streamers with our synthetic images by calculating a synthetic ``monthly'' background in a way similar to the procedure done for real line-of-sight images. From our steady-state, we produce 27 synthetic images by positioning satellites separated by increments of 13$^\circ$ in the ecliptic plane, to simulate the effects of a full solar rotation, each image representing LASCO approximate view on a different day of the Carrington rotation;  the background image is calculated from this by taking, for each pixel, the minimum value of the 27 images. 
Comparing the first and third columns of Figure \ref{fig:LASCO}, one can see that the streamer around PA 120 is still significantly deflected in front of the second CME as compared to its pre-eruption value. The ``flattening'' and associated ``dimple'' of the second CME front around PA 40, which can be seen in the real and synthetic images is due to the second CME propagating into one of these deflected streamers.  

\begin{figure}[ht] 
 \centerline{\includegraphics*[width=12cm]{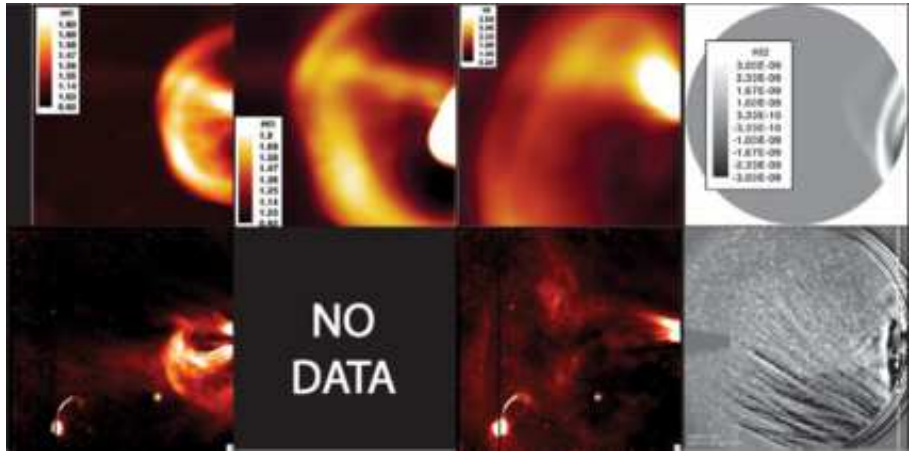}}
 \caption{Comparison of synthetic ({\it top}) and real ({\it bottom}) SECCHI/HI images. The first three columns are HI-1-A images (background-divided for the synthetic images), the last one HI-2 images (2-hr running difference). The times are January 25 02:00UT, 10:00UT, and January 26 00:00UT, 04:00UT from left to right. There was no SECCHI observations between 06:00UT and 23:00UT on January 25. The real HI-1 images are processed by the NRGF.}
 \label{fig:HIs}
\end{figure}

In Figure~\ref{fig:HIs}, we compare synthetic images of the two CMEs in the heliosphere from the point-of-view of STEREO-A with real images when available. There is also an online animation combining the synthetic line-of-sight images from LASCO and the HIs into one continuous movie covering over 3 days. 
There is a fair agreement between the overall shape and position of the CME in real and synthetic images of the first CME (left column of Figure~\ref{fig:HIs}). However, the CME in the synthetic image extends significantly north of the bright feature corresponding to the north-east streamer. This might be due to an over-expansion of the flux rope CME due to the unrealistic CME initiation mechanism. The CME also appears significantly brighter in the synthetic image compared to the real one: this might be partly explained by the difference in the background subtraction mechanism. Next, we show an example of a HI-1 image during SECCHI downtime corresponding to the top left image in Figure~\ref{fig:2D} and Figure~\ref{fig:3D} -- i.e., at the start of the interaction between the CMEs --. This demonstrates that HI-1 would have been able to clearly observe the two CME fronts at the same time, allowing for a more direct comparison of the CME evolution in the heliosphere. We believe that the image at 00:01UT on January 26, once processed by the Normalized Radial Graded Filter (NRGF, see \opencite{Morgan:2006}) shows evidence of the second CME front exiting the instrument field-of-view. This is confirmed by an analysis of a time-sequence of our synthetic images which show the second front propagating inside the first one starting at 18:00UT on January 25 (see animation appended to the electronic version). 
Last, we show one example of an HI-2 image at the same time as the bottom right image in Figure~\ref{fig:2D}. One has to remember that the plane in Figure~\ref{fig:2D} corresponds to PA 90 which is rolled counter-clockwise by approximately 21.7$^\circ$ in the images in Figure~\ref{fig:HIs}. As analyzed in \inlinecite{Lugaz:2008b}, the brightest front corresponds to the merged CMEs while the very dim leading front corresponding to part of the first front along PAs close to the Thomson sphere (TS) which have not yet been overtaken by the second CME. 

\subsection{Time-Elongation Maps}
\begin{figure}[ht*] 
 \centerline{\includegraphics*[width=7.5cm]{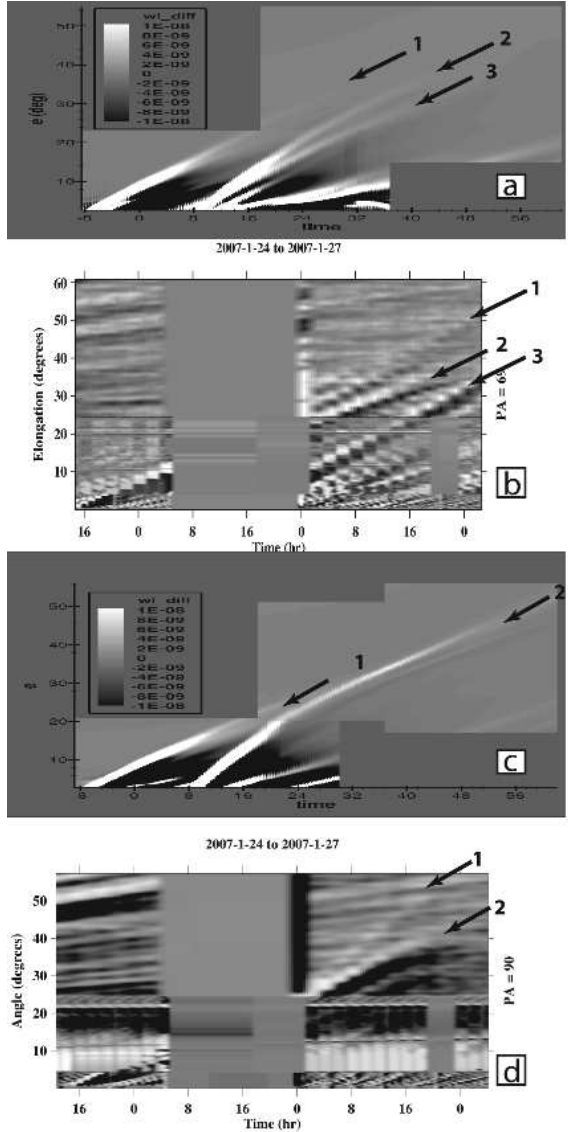}}
 \caption{Comparison of synthetic (a and c) and real (b and d) J-maps for PA~69 (a and b) and PA~90 (c and d). PA~69 is the center of the HIs image while PA~90 is closer to the ``nose'' of the CMEs. Time starts at 16:00UT on January 24 (corresponding to -8 on the panels a and c and the first 16 on panels b and d).}
 \label{fig:Jmap}
\end{figure}

Since qualitative comparisons between synthetic and real white-light observations show that the simulation offers an accurate enough reproduction of the observations, it is now important to compare more quantitively simulated and real images. One of the methods to study the evolution of density enhancements
is to produce time-elongation maps (J-maps) for different PAs \cite{Sheeley:1997, Sheeley:2008a, Rouillard:2008, Davies:2009}. Such maps allow for the tracking of CMEs to large elongation angles and the study of their evolution without assumptions concerning the direction of propagation. In this section, we discuss J-maps along two PAs: 90 and 69, corresponding to the approximate ``nose'' of the CMEs and the apparent central angle of the SECCHI instruments, respectively. Synthetic and real maps are shown in Figure~\ref{fig:Jmap}; we first identify the tracks seen in these maps before comparing them quantitatively. 

For the synthetic J-map, we take a slice every 20  minutes and plot the difference of the total brightness with the slice 2 hours earlier. 
For the real J-map, a slice is taken for every observation (i.e. every 2 hours for HI-2 and varying between every 20 minutes and every hour for HI-1) and the running difference is plotted. The higher cadence makes the synthetic J-maps look ``smoother'', but because we take a 2-hr running difference, the width and slope of the tracks have the same meaning as for the real maps. SECCHI instruments did not image from 04:00UT on January 25 to 00:00UT on January 26. The presence of Venus make the HI-1 data near PA~90 difficult to use. 

In the synthetic J-map at PA~69, there are three main distinct tracks (marked with black arrows). They correspond to the three fronts described in \inlinecite{Lugaz:2008b}: the first ejection, the second ejection and a dense stream (tracks 1 to 3, respectively). The dense stream is first compressed by the first CME and can be seen around 08:00UT on January 25 behind the first CME. The compressed (resp. uncompressed) stream can be seen ahead of (resp. behind) the first CME in the {\it top left panel} of Figure \ref{fig:2D}. At 04:00UT on January 26, there are three fronts past 20$^\circ$ elongation (i.e. in HI-2 field-of-view), the brightest corresponding to the second CME. Noteworthy is the fact that the 2 CMEs do not appear to merge along this PA, and that the track of the first CME becomes very faint past 30$^\circ$ elongation. 

The real J-map at this PA is much more complicated and harder to analyze in part due to the lack of observations on January 25. There also appears to be three main tracks visible in HI-2 starting at about 14:00UT on January 26. We believe that these three tracks correspond to the fronts identified in the synthetic image in the same order. One apparent problem with this interpretation is that track 1 has the fastest angular speed of all and can be tracked the farthest until about 60$^\circ$ elongation, whereas track 2 (corresponding to the second eruption, in this interpretation) is the brightest but fades away quickly past 35$^\circ$ elongation. According to our analysis, track 1 initially (in HI-1) corresponds to the first CME and at later times (in HI-2) to the merged CMEs, which explains the faster angular speed there. This track corresponds to the ``forerunner'' structures described by \inlinecite{Harrison:2008}.
Making a definitive identification regarding the origin of track 1 would require to know if tracks 1 and 2 merge during SECCHI downtime, since the explanation described above would require the two tracks to merge when the two CME fronts collide. If this analysis is correct, there is one important aspect that the simulation does not capture: the fading of the second (and brightest) front. 
The most likely explanation is that this front and its brightening are associated with a temporary phenomenon: the propagation of the second CME inside the first dense sheath, resulting in twice compressed medium. As the CMEs merge, front 2 corresponds to the density peak of the twice compressed sheath, whereas front 1 corresponds to the new sheath associated with the merged CMEs, similar to what is explained in \inlinecite{Lugaz:2005b}. However, the second density peak eventually relaxes, resulting in the rapid fading of track 2. This phenomenon does not happen in the synthetic line-of-sight images, but  it would happen if the merging happened approximatively at the same time along all directions. In our simulation, as described in the next section and as seen in Figure \ref{fig:2D}, the two CMEs do not merge until 02:00UT on January 27 along the TS.

In the synthetic J-map at PA~90, there are only two main distinct tracks, corresponding to the two ejections. The dense stream (which corresponds to track 3 at PA~69) does not appear at this PA, reflecting the three-dimensional nature of the heliosphere and giving us an additional clue that this feature is not due to a CME (as explained in \citeauthor{Lugaz:2008b}2008). Along this angle, there is complete merging of the fronts around 06:00UT on January 26. The merging is associated with a brightening of the front, especially noticeable between 08:00UT and 18:00UT on January 26. The real J-map also shows only two tracks with the second one brighter, but they appear to diverge from a similar (merged) position at the start of the observations on January 26. 
%%%%%%%%%%%%%%%%%%%%%%%%%%
\begin{figure}[ht] 
\begin{minipage}[]{1.0\linewidth}
\begin{center}
{\includegraphics*[width=6.cm]{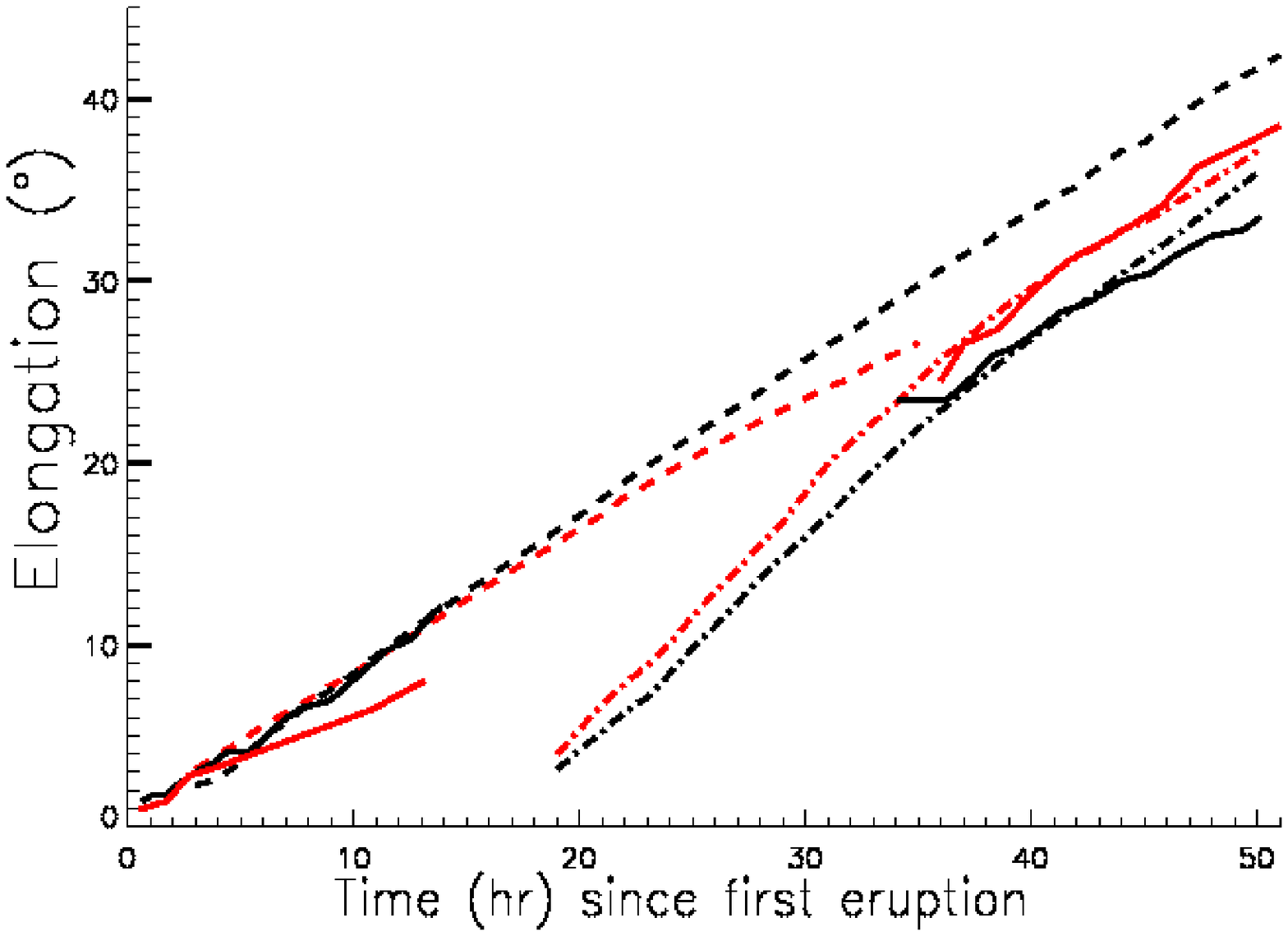}} 
{\includegraphics*[width=6.cm]{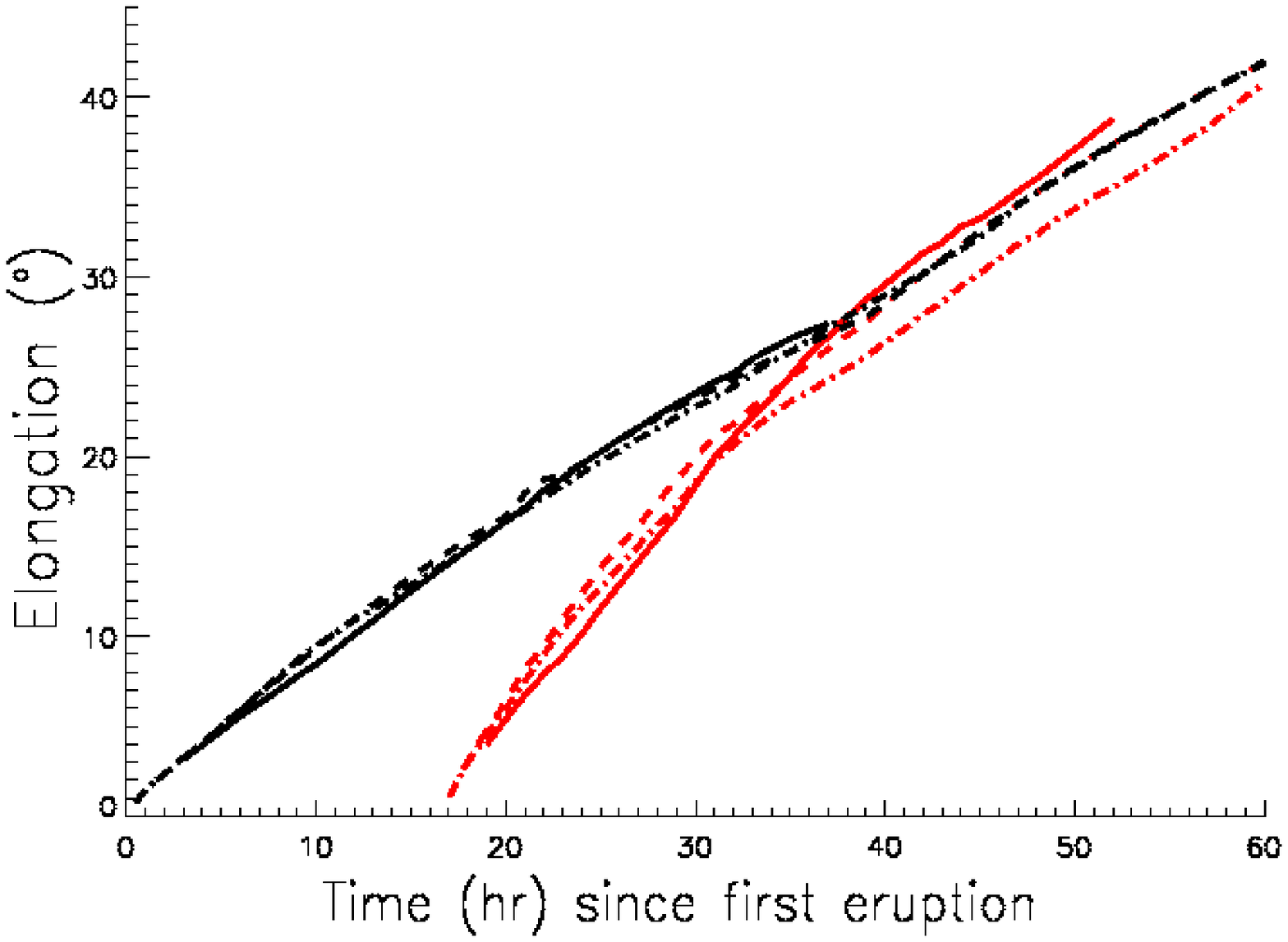}}
\end{center}
\end{minipage}\hfill
 \caption{Comparison between time-elongation profiles from the observations, synthetic images and derived from the 3-D simulation. The left panel shows a comparison of the real ({\it solid}) and synthetic ({\it dash}: first eruption; {\it dash-dot}: second eruption) elongation angles for PA~69 (red) and PA~90 (black). The right panel shows a comparison of the synthetic elongation ({\it solid}), elongation from the simulation on the limb ({\it dash}) and on the  Thomson sphere ({\it dash-dot}) for PA~90 for the first (black) and second (red) eruptions.}
 \label{fig:1D}
\end{figure}
%%%%%%%%%%%%%%%%%%%%%%%%%
Similarly to the J-map at PA~69, we believe  track 2 is associated with the merging of the ejections (propagation of the second front inside the first one) while track 1 is associated with the first ejection and later on with the merged ejections. In our synthetic images, there is also a splitting of the merged front along PA~90 starting around 02:00UT on January 27, as can be seen in the online animation. This corresponds more or less to the time of the merging of the CME fronts along the TS in the simulation.

\subsection{Time-Elongation and Time-Height Plots}

Next, we focus on the quantitative comparison between observed and synthetic positions from the J-maps, focusing on the two CMEs (tracks 1 and 2 according to the analysis above). 
The left panel of Figure \ref{fig:1D} confirms without doubt that the brighter front observed in HI-2 corresponds to the second eruption (PA~69) or the merged eruptions (PA~90). The model is in good agreement with observations except at large elongation angles (past 30$^\circ$), starting approximatively 45 hours after the first eruption. This might be due, for example, to the complexity of tracking dim fronts in the real images, or to some observational effects not being well reproduced by the synthetic line-of-sight procedure. We can now compare the position derived from the line-of-sight images with that from our 3-D simulation to determine which parts of the CME is being tracked. 

In the right panel of Figure \ref{fig:1D}, we show this comparison for PA~90 (the ``nose'' of the ejections). The solid, dash and dash-dotted lines show the elongation angles of the CME fronts in the synthetic line-of-sight, 90$^\circ$ from STEREO-A and on the TS, respectively. The TS is the locus of maximum scattering \cite{Jackson:1985, Vourlidas:2006}. It comes from the fact that the scattering along a given line of sight is maximized at the point of closest approach to the Sun.  Since the source region of the CMEs was behind the eastern limb of the Sun, assuming that the  CMEs are propagating along the limb (90$^\circ$ from STEREO) is the best approximation which can be made. By analyzing this figure, we can see that, for the first eruption, the line-of-sight position is always within 1$^\circ$ of the real one. The position can, then, be estimated with minimal error by $R_{\mathrm{CME}} = 
d_{\mathrm{STEREO}} \tan(\epsilon)$, where $d_{\mathrm{STEREO}}$ = 0.97~AU is the heliospheric distance of STEREO-A, $\epsilon$ is the elongation angle and $R_{\mathrm{CME}}$ is the position of the CME 90$^\circ$ away from STEREO-A. From this position, the speed and heliospheric deceleration of the first CME can then be relatively well constrained up to about 90~$R_{\odot}$.

For the second CME, we must distinguish the periods before and after the CME-CME interaction. Before the interaction (until about 30 hours), assuming that the observed position corresponds to the position on the TS is  a better approximation than the ``limb'' approximation. For example, at time t = 28 h, the position on the TS is 16.2$^\circ$, on the limb (90$^\circ$ from STEREO-A ) it is 17.5$^\circ$ and from the synthetic line-of-sight, it is 15.5$^\circ$ (as seen in the right panel of Figure~\ref{fig:1D}). 
%Relying on line-of-sight images to determine the exact time of the merging of the two CME fronts in three-dimension can be a tricky task. The CME fronts appear to merge 38 hours after the launch of the first CME on the synthetic line-of-sight images, which correspond approximatively to the time of the merging along the eastern limb in our simulation (after 40 hours). Yet, on the Thomson sphere, they have not yet merged on after 60 hours. This is because as the elongation angle increases, the TS intercepts more and more of the ``flanks'' of the CMEs where the merging happens later (see Figure ~\ref{fig:2D}).

At large elongation angles, one must also consider the curvature of the CMEs and the effect of the TS to determine the radial distances. 
Assuming a propagation along the limb (or any fixed angular position) may result in a large overestimation of the position. 
For example, at time $t$ = 52 hours (18:00UT on January 26), the dim front along PA~69 is at 43.3$^\circ$, which is similar to the ``forerunner'' front at 42$^\circ$ described by \inlinecite{Harrison:2008}. 
This angular position corresponds to radial positions of 145~$R_{\odot}$ assuming that the emission originates from the TS, yet 200~$R_{\odot}$ assuming it originates from the limb (90$^\circ$ east). In fact, based on our simulation, the front of the first CME was at about 145~$R_{\odot}$ on the TS but only 160~$R_{\odot}$ at 90$^\circ$. 

However, knowing the de-projected distances along the TS does not allow for the determination of the true CME speed. This is because, at each time instant, the position on the TS corresponds to a different part of the CME and a time-height profile derived from these positions is not consistent with tracking the same part of the CME along a fixed direction. Therefore such a profile would show an apparent acceleration or deceleration of the CME based on its interaction with the TS. Contrary to LASCO observations where the bright fronts were tracked for hours, tracking density enhancements for days with SECCHI does not allow us to use a simple approximation such as the ``plane-of-sky'' one. 
Therefore, the best way to use the information provided by the line-of-sight images is to assume that the CME is spherically symmetric. Then, the position on the TS can be assumed to correspond to the position along any direction. For the example above, this is a better approximation that assuming the bright front is at 90$^\circ$ (15~$R_{\odot}$ underestimation vs. 40~$R_{\odot}$ overestimation of the actual position). Other spurious effects, such as the appearance of multiple bright fronts in the line-of-sight images associated with one CME front are expected to appear at very large elongation angles \cite{Manchester:2008}.

\section{Discussion and Conclusion} \label{sec:Conclusion}

We have investigated the heliospheric evolution and interaction of the two coronal mass ejections in January 24-27, 2007, using a numerical MHD simulation to interpret and better analyze observations by the SECCHI suite aboard STEREO. We are able to reproduce successfully the observations in LASCO field-of-view. Using a realistic background subtraction procedure for the synthetic line-of-sight images, we are able to reproduce observations of streamer deflections and to relate the position of the new streamer with a deformation and ``flattening'' of the second CME along some position angles. 

We have proven that STEREO has the ability to make indisputable observations of a CME catching up with a previous one, which would help constraining numerical models of CME-CME interaction. Assuming no deceleration of the two CMEs and using the time-height data from LASCO, the merging of the two CMEs was expected to happen around 21:00UT on January 25 around 25$^\circ$ elongation. We, in fact, find evidence that the CMEs had just merged at the beginning of January 26 when SECCHI started observing again. In our simulation, the merging happens around 06:00UT on January 26. We believe that the bright front around elongation 20-24$^\circ$ in the HI-1 image at 00:00UT and 02:00UT on January 26 corresponds to the front of the second CME propagating inside the front of the first CME or a second density peak associated with twice-shocked medium. There were also observations by SMEI  of one bright front from the end of January 25 to about 08:00UT on January 26 whose position concurs with that of the second front (brightest) in our J-maps (D. Webb, private communication). We believe another front, which can be seen in the SMEI images from 12:00UT on January 27 to 06:00UT on January 28, mostly at large PAs (up to 120$^\circ$) corresponds to the front 1 in the J-maps. Merging of the SMEI and SECCHI data will probably help tracking density structures in the heliosphere, but careful cross-calibration of the two instruments is required first. 
 
Finally, our simulation helps us identify the different tracks in time-elongation maps (J-maps) of the CMEs and relate the elongation angles to the actual positions of the CME fronts. We find tracks associated with the two CME fronts, the merged front and a dense stream. The time-elongation profile of the brightest front associated with the CMEs is in good agreement qualitatively and quantitatively with J-maps constructed from observations. 
However, there is one important aspect where our model differs from observations: in the real images, the brightest front, which corresponds initially to the second CME, rapidly fades after the CME-CME merging, whereas in our synthetic images, the initially dimmest front, corresponding to the first CME, is the one which fades quickly. One possible explanation is that the second front after the CME-CME merging, in the real images, is associated with the presence of a second density peak in the merged sheath associated with the twice-compressed medium (as discussed in \opencite{Lugaz:2005b}). Our simulation may not capture this in the line-of-sight images for a number of reasons, including limited spatial resolution, asymmetry of the second eruption and too late merging of the CME fronts.

\inlinecite{Rouillard:2008} associated a bright front observed in J-maps with in-situ observations of a corotating interaction region at L1. Here, it is not possible to make such association, because the CMEs did not hit Earth. However, the simulation allows us to fill SECCHI data gap and to explain the difference in the number of tracks in different PAs. We found that one of the observed fronts can be tracked until about 60$^\circ$ which corresponds to 0.85~AU on the Thomson sphere and about 1.75~AU if one assumes a point source propagating 90$^\circ$ away from Earth. The latter position is not consistent with the CME speed and time evolution (barring dramatic late acceleration). Therefore, it proves that HI-2 (and SMEI at large elongation angles) essentially captures the propagation of the CMEs onto the Thomson sphere surface. This behavior was predicted by \inlinecite{Vourlidas:2006}. This fact should make the direct use of Heliospheric Imagers data to derive a ``true'' speed of CMEs  or to test heliospheric drag models delicate. The procedure followed here is to use a numerical simulation with synthetic imaging capabilities to derive the likely speed, deceleration and interaction of the CMEs. The first step is to make sure the synthetic observations reproduce successfully the real observations, before determining the CME position in 3-D. We also show that, in the absence of 3-D numerical models %(which can not yet be used for every CME observations by SECCHI) 
and triangulation of the CMEs by the two STEREO spacecraft, assuming spherical symmetry of the CME 
and deriving its position from the projection on the Thomson sphere is the most accurate way to determine the best approximation of the ``true'' position of the CME front. It is also possible that a number of case studies of events observed by SECCHI/HIs will help to develop a paradigm to obtain time-height profiles from heliospheric imagers, using for example 3-D reconstruction of the CMEs such as done by \inlinecite{THoward:2008b}.

%% Acknowledgements
%
 \begin{acks}
The research for this manuscript was supported by NSF grants ATM-0639335 and ATM-0819653 as well as NASA grants
  NNX-07AC13G and NNX-08AQ16G. N.~L. would also like to acknowledge travel support from SPS to participate to the European Solar
  Physics Meeting in Freiburg to present these results and thank D.~Webb and V.~Bothmer for discussion about these eruptions at the IAU 257 Symposium and EGU meeting respectively.
  The simulation reported here was carried out on a dedicated cluster of the solar group at the Institute for Astronomy.
  We would like to thank the reviewer and the editors for helping us to improve and to clarify this manuscript.
  SOLIS data were obtained from the NSO website.
   The SECCHI data are produced by an international consortium of 
  Naval Research Laboratory, Lockheed
  Martin Solar and Astrophysics Lab, and NASA Goddard Space Flight
  Center (USA), Rutherford Appleton Laboratory, and University of
  Birmingham (UK), Max-Planck-Institut f{\"u}r Sonnensystemforschung
  (Germany), Centre Spatiale de Liege (Belgium), Institut d'Optique
  Th{\'e}orique et Appliqu{\'e}e, and Institut d'Astrophysique
  Spatiale (France).  SoHO is a project of international cooperation between ESA and NASA,  and the SOHO LASCO/EIT catalogs are  maintained by NASA, the Catholic University of America, and the US Naval Research Laboratory (NRL).
 %SMEI data are produced by a team from the US Air Force Research Laboratory, the University of Birmingham in the United Kingdom, the University of California at San Diego, Boston College, and Boston University, and maintained by the Air Force Research Laboratory . 
 \end{acks}

%%% %%%%%%%%%%%%%%%%%%%%%%%%%%%%%%%%%%%%%%%%%%%%%%%%%%%%%%%%%%%
%% Bibliography
\bibliographystyle{spr-mp-sola-cnd}
%\bibliography{thesis}

\end{article} 
\end{document}